\documentclass[journal]{IEEEtran}

\usepackage{amsmath,amssymb,amsfonts}
\usepackage{graphicx}
\usepackage{textcomp}
\usepackage{xcolor}

\def\BibTeX{{\rm B\kern-.05em{\sc i\kern-.025em b}\kern-.08em
    T\kern-.1667em\lower.7ex\hbox{E}\kern-.125emX}}
\usepackage{xcolor}
\usepackage[linesnumbered,ruled,vlined]{algorithm2e}
\usepackage{listings}
\usepackage[noend]{algpseudocode}
\SetCommentSty{mycommfont}
\SetKwInput{KwInput}{Input}                % Set the Input
\SetKwInput{KwOutput}{Output}              % set the Output
\usepackage{subcaption}

\usepackage{balance}
\usepackage{cite}
\usepackage{soul}
\usepackage{multirow} % for multi row in tables
\usepackage{acro}
\DeclareMathOperator{\EX}{\mathbb{E}}% expected value

\DeclareAcronym{CIR}{
  short = CIR,
  long  = channel impulse response
}
\DeclareAcronym{CFR}{
  short = CFR,
  long  = channel frequency response
}
\DeclareAcronym{OFDM}{
  short = OFDM,
  long  = orthogonal frequency division multiplexing
}
\DeclareAcronym{THz}{
  short = THz,
  long  = terahertz
}
\DeclareAcronym{PS}{
  short = PS,
  long  = phase shifter
}
\DeclareAcronym{PL}{
  short = PL,
  long  = path loss
}
\DeclareAcronym{LAA}{
  short = LAA,
  long  = lens antenna array
}
\DeclareAcronym{LAS}{
  short = LAS,
  long  = lens antenna subarray
}
\DeclareAcronym{SNR}{
  short = SNR,
  long  = signal-to-noise ratio
}
\DeclareAcronym{LoS}{
  short = LoS,
  long  = line-of-sight
}
\DeclareAcronym{RF}{
  short = RF,
  long  = radio-frequency
}
\DeclareAcronym{JRC}{
  short = JRC,
  long  = joint radar and communication
}
\DeclareAcronym{mmWave}{
  short = mmWave,
  long  = millimeter-wave
}
\DeclareAcronym{AoA}{
  short = AoA ,
  long  = angle of arrival
}
\DeclareAcronym{AoD}{
  short = AoD,
  long  = angle of departure
}
\DeclareAcronym{MIMO}{
  short = MIMO,
  long  = multiple-input multiple-output
}
\DeclareAcronym{mMIMO}{
  short = mMIMO,
  long  = massive multiple-input multiple-output
}
\DeclareAcronym{UWB}{
  short = UWB,
  long  = ultra wideband
}
\DeclareAcronym{BS}{
  short = BS,
  long  = base station
}
\DeclareAcronym{UE}{
  short = UE ,
  long  = user equipment
}
\DeclareAcronym{ISI}{
  short = ISI,
  long  = inter-symbol interference
}
\DeclareAcronym{ULA}{
  short = ULA,
  long  = uniform linear array
}
\DeclareAcronym{SPMT}{
  short = SPMT,
  long  = single pole multiple throw
}
\DeclareAcronym{MPMT}{
  short = MPMT,
  long  = multiple pole multiple throw
}
\DeclareAcronym{FFT}{
  short = FFT,
  long  = fast Fourier transform
}
\DeclareAcronym{DAC}{
  short = DAC,
  long  = digital-to-analog converters
}
\DeclareAcronym{EE}{
  short = EE,
  long  = energy efficiency
}
\DeclareAcronym{PCB}{
  short = PCB,
  long  = printed circuit board
} 
\DeclareAcronym{SAW}{
  short = SAW,
  long  = surface acoustic wave
} 
\DeclareAcronym{BAW}{
  short = BAW,
  long  = bulk acoustic wave
}
\DeclareAcronym{CMOS}{
  short = CMOS,
  long  = complementary metal-oxide semiconductor
}
\DeclareAcronym{QFN}{
  short = QFN,
  long  = quad flatpack non-lead
}
\DeclareAcronym{mMTC}{
  short = mMTC,
  long  = massive machine type communication
}

\DeclareAcronym{IoT}{
  short = IoT,
  long  = Internet-of-Things
}

\usepackage[compact]{titlesec}
\titlespacing{\section}{1.0pt}{*1.0}{*0}
\titlespacing{\subsection}{1.1pt}{*1.1}{*0}
\titlespacing{\subsubsection}{0.3pt}{*0}{*0}

\SetAlCapNameFnt{\scriptsize}
\SetAlCapFnt{\scriptsize}
\SetAlFnt{\small}

\begin{document}

\title{Novel Transceiver Design in Wideband Massive MIMO for Beam Squint Minimization}

\author{
    \IEEEauthorblockN{Liza Afeef,}
    \IEEEauthorblockN{Abuu B. Kihero,} 
	\IEEEauthorblockN{and}
	\IEEEauthorblockN{H\"{u}seyin Arslan,}
	\IEEEmembership{Fellow, IEEE}

\thanks{This work was supported in part by the U.S. National Science Foundation under Award ECCS-1923857. The authors are with the Department of Electrical and Electronics Engineering, Istanbul Medipol University, Istanbul, 34810, Turkey (e-mail: liza.shehab@std.medipol.edu.tr; abuu.kihero@std.medipol.edu.tr; huseyinarslan@medipol.edu.tr). H. Arslan is also with Department of Electrical Engineering, University of South Florida, Tampa, FL, 33620, USA.}
\thanks{This work has been submitted to the IEEE for possible publication. Copyright may be transferred without notice, after which this version may no longer be accessible.}
}

\maketitle

\begin {abstract}
When using ultra-wideband (UWB) signaling on massive multiple-input multiple-output (mMIMO) systems, the electromagnetic wave at each array element incurs an extra propagation delay comparable to (or larger than) the symbol duration, producing a shift in beam direction known as beam squint. The beam squinting problem degrades the array gain and reduces the system capacity. This letter proposes a novel transceiver design based on lens antenna subarray (LAS) and analog subband filters to compensate for the beam squinting effect. In particular, the proposed design aims to divide the UWB signal into narrowband beams and control them with a simplified exhaustive search-based precoding that is proposed to align the beam angle to the target direction. The design is analyzed in terms of beam gain, complexity, power consumption, and capacity, demonstrating significant performance enhancement with respect to the conventional system with uncompensated beam squinting problem.

\end {abstract}


\IEEEpeerreviewmaketitle

\section {Introduction} \label{sec:introduction}

\Ac{mMIMO} and \ac{UWB} transmission have been considered among the candidate technological enablers for enhancing the performance and efficiency of the next-generation wireless networks. \ac{mMIMO} can improve spectrum and energy efficiency, combat small scale fading through channel hardening, and extend network coverage by overcoming the \ac{PL} problem \cite{busari2017millimeter}. \Ac{UWB} transmission not only facilitates high data rate communication but also provides resilience against multipath fading and covertness against jamming attacks \cite{cassioli2002ultra}. Advantages of both \ac{mMIMO} and \ac{UWB} also span to the \ac{RF} sensing aspect of the wireless networks that have been recently considered under the \ac{JRC} framework \cite{liu2020joint}. They, respectively, provide fine spatial and temporal multipath resolution, thereby facilitating accurate localization and ranging of the target objects. 

Despite these desirable advantages, it has been shown that \ac{mMIMO} systems implementing \ac{UWB} signaling suffer from \textit{spatial-wideband effect}, recapitulated hereunder. The Huygens-Fresnel wave propagation principle dictates that, in the antenna array systems, unless the incident signal is perpendicular to the array, the received signal at different array elements is a slightly delayed version of the original signal. The amount of delay incurred across the elements depends on the inter-element spacing and the signal's \ac{AoA}/\ac{AoD}. For a system with a relatively small number of antennas, as it is in the conventional small-scale \ac{MIMO} systems, the maximum delay across the antenna aperture can be much smaller than the symbol duration, and thus its effect can be ignored. However, with high-dimensional antenna arrays, i.e., \ac{mMIMO}, with \ac{UWB} signaling, this delay can be in the order or even larger than the symbol duration, leading to \textit{delay squinting effect} in the spatial-delay domain, i.e., significant delay spread is observed across the array even in the pure \ac{LoS} propagation condition. The delay squinting problem renders the steering vector frequency-dependent in the angular-frequency domain. That is, in multicarrier systems like \ac{OFDM}, signals at different subcarriers will point to different physical directions. Signals pertained to such derailed subcarriers might not arrive at the intended receiver or align with sidelobes or nulls of the receiver's radiation pattern, thereby degrading the performance. This phenomenon is referred to as \textit{beam squinting} \cite{wang2018spatial_2}.

Numerous approaches have been proposed in the literature to tackle the beam squinting problem. The works in \cite{liu2013minimize,cai2016effect,letter2021} approach the problem from codebook design perspective. Although some promising results have been reported in these studies, complexity issues, stemming from hardware requirements or prohibitively large codebook sizes, seem to haunt the proposed approaches. Authors of \cite{liu2018space} proposed an Alamouti-based beamforming scheme that minimizes beam gain variation of all subcarriers within the operational bandwidth. Since the proposed beam pattern optimization involves Eigenvalues and vectors computation, the computation complexity of this scheme grows with the number of antennas. Hybrid precoders designed to deal with the beam squint in the digital domains are also proposed in \cite{2019beam} and \cite{li2018beam}. However, the improvement gained through such digital baseband processing is minimal. In \cite{sattar2019antenna}, analog architecture design that employs bandpass filters and extra \acp{PS} to facilitate a subband-based beam squinting compensation is presented. The usage of extra \acp{PS} in this work has an obvious drawback of the reduced energy efficiency.

\begin{figure}[t!]
    \begin{center}
    \includegraphics[scale=0.5]{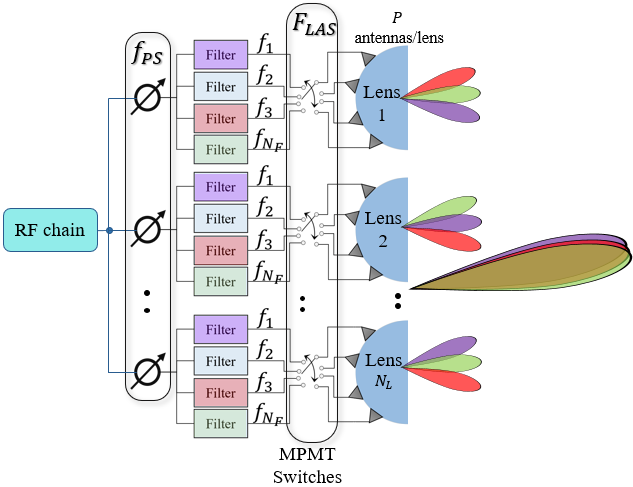}
    \end{center}
    \centering
    \caption{The proposed LAS analog transceiver design for a multicarrier signal in mMIMO systems.}
\label{fig:LAS_system}
\vspace{-17pt}
\end{figure}

In this work, a modified \ac{LAS} analog transceiver design is proposed to address the beam squinting problem in \ac{UWB} \ac{mMIMO} systems. The \ac{LAS} design involves the use of both \acp{PS} and switching networks to steer the beam \cite{Murat,karabacak2020hybrid}. Unlike the traditional \ac{LAS} systems which mainly focus on increasing the field of view and steering resolution, this work leverages the combination of these two steering mechanisms (i.e., \acp{PS} and switches) to compensate for the beam squinting problem while maintaining the intended beamwidth performance. Mainly, \acp{PS} are used to steer the beam to the desired direction based on the location of the targeted user, and the switching mechanism is used to select an antenna element under the lens that can correct/minimize the deviation of the beam (due to squinting) from the intended direction. Considering the fact that beam squinting increases with the subcarrier's frequency with respect to the system's center frequency, in the proposed design, the wideband signal is chunked into several subbands in which the subcarriers would be derailed by relatively the same amount. This is achieved by inserting a bank of analog subband filters between \acp{PS} and the lenses, as depicted in Fig. \ref{fig:LAS_system}. In order to correct the expected beam squinting for each subband, a low-complex simplified search-based precoding is proposed to select the appropriate antenna element under the lenses.
The main contributions of this letter are summarized below:
\begin{itemize}
    \item A modified \ac{LAS} design for mitigating beam squinting problem is presented and evaluated. Unlike the contemporary approaches presented in \cite{liu2013minimize} and \cite{sattar2019antenna} that employ analog filters and extra \acp{PS} in their designs, our proposed design inherits the energy efficiency nature of the \ac{LAS} structure, making it more appealing not only from affordability perspective but also for the envisioned green-communication networks.
    \item While the proposed design can work with the traditional exhaustive search-based precoder for antenna selection under the lenses, an enhanced, threshold-based precoder is proposed to reduce the search complexity without notable degradation on the beam-gain performance.
\end{itemize}

The rest of this letter is organized as follows. The system model of the conventional \ac{mMIMO} that leads to the beam squinting problem is recapped in Section II. Detailed explanation and analysis of the proposed \ac{LAS} design and its proposed precoder are given in Section III, followed by the numerical evaluation in Section IV. Section V finally concludes the work.

\section{System Model and Problem Formulation}
A \ac{mMIMO} system with a \ac{BS} equipped with $M$-antennas \ac{ULA} is considered. Suppose there are $L_p$ channel paths arriving at the \ac{BS} where each path $\ell \in \{1, 2, \dots, L_p\}$ is associated with a passband complex gain $\alpha_{\ell}$, \ac{AoA}/\ac{AoD} $\hat{\theta}_{\ell}\in[-\pi/2,\pi/2)$, and the propagation delay $\tau_{\ell}\in[0,\tau_{\text{max}}]$, where $\tau_{\text{max}}$ is the channel's maximum delay spread. As briefly mentioned in Section \ref{sec:introduction}, for a \ac{mMIMO} system implementing a very wideband signaling, there exists a non-trivial amount of delay across the array elements (with respect to the first element), given by \cite{wang2018spatial_2}
\begin{equation}
 \Delta\tau_{m,\ell} = (m-1)\frac{d\sin\hat{\theta}_{\ell}}{c}
\label{eq:squintDelay}
\end{equation}
with $d=\lambda_c/2$ being inter-element spacing where $\lambda_c$ is the carrier wavelength, and $m\in\{1,2,\dots, M\}$. Consequently, the total delay of an $\ell^{th}$ path observed by $m^{th}$ element can be given as 
\begin{equation}
    \tau_{m,\ell} = \tau_{\ell} + \Delta\tau_{m,\ell}.
\end{equation}
Accordingly, the \ac{CIR} observed by the $m^{th}$ element is given as 
\begin{equation}
    h_{m}(t) = \sum^{L_p}_{\ell=1}\alpha_{\ell} e^{-j2\pi f_c\tau_{m,\ell}}\delta(t-\tau_{m,\ell}),
    \label{eq:CIRequation}
\end{equation}
where $f_c$ is the carrier frequency. 
% ------------------ figure ------------------------- %
% --------------------------------------------------- 
Taking the Fourier transform of \eqref{eq:CIRequation} and simplifying, the \ac{CFR} of the considered link is obtained as \cite{wang2018spatial_2}
\begin{equation}
    h_{m}(f) = \sum^{L_p}_{\ell=1}\tilde\alpha_{\ell}e^{-j2\pi(m-1)\frac{d\sin\hat{\theta}_{\ell}}{\lambda_c}}\underbrace{e^{-j2\pi(m-1)\frac{fd\sin\hat{\theta}_{\ell}}{c}}}_{\text{squint-inducing term}}e^{-j2\pi f\tau_{\ell}},
\end{equation}
where $\tilde\alpha_{\ell} = \alpha_{\ell}e^{-j2\pi f_c\tau_{\ell}}$ is the equivalent baseband path gain. Stacking the channels from all $M$ antennas, the \ac{CFR} vector over the whole array can be expressed as
\begin{equation}
    \mathbf{h}(f) = \sum^{L_p}_{\ell=1}\tilde\alpha_{\ell}\underbrace{\mathbf{a}_{\text{ideal}}(\hat{\theta}_{\ell})\odot\mathbf{a}_{\text{squint}}(\hat{\theta}_{\ell},f)}_{=~\mathbf{a}(\hat{\theta}_{\ell},f)}e^{-j2\pi f\tau_{\ell}},
    \label{eq:CFR ideal squint}
\end{equation}
where $\odot$ represents the Hadamard product. $\mathbf{a}_{\text{ideal}}(\hat{\theta}_{\ell}) = [1, e^{-j2\pi\frac{d\sin\hat{\theta}_{\ell}}{\lambda_c}}, \dots, e^{-j2\pi (M-1)\frac{d\sin\hat{\theta}_{\ell}}{\lambda_c}}]$ is the perfect spatial-domain steering vector and $\mathbf{a}_{\text{squint}}(\hat{\theta}_{\ell},f) = [1, e^{-j2\pi\frac{fd\sin\hat{\theta}_{\ell}}{c}}, \dots, e^{-j2\pi (M-1)\frac{fd\sin\hat{\theta}_{\ell}}{c}}]$ is a frequency-dependent vector that induces beam-squinting problem in the system. As such, $\mathbf{a}_{\text{squint}}$ modifies $\mathbf{a}_{\text{ideal}}$ into a frequency-dependent effective steering vector $\mathbf{a}$. In the multicarrier signaling, $\mathbf{a}$, being frequency-dependent, directs signal pertaining to different subcarriers to different directions away from the desired direction. From \eqref{eq:CFR ideal squint}, the effective direction of the signal at subcarrier $k$ centered at $f_k$ $=((k-1)-(N-1)/2)\times \Delta f + f_c$, where $k=1,2,\cdots,N$ with $N$ and $\Delta f$ being the \ac{FFT} size and subcarrier spacing, respectively, can be found as  % \cite{wang2018digital}
\begin{equation}
    \theta (f_k) = \sin^{-1} \left( \frac{\sin \hat{\theta}_\ell}{1+\frac{f_k}{f_c}} \right).
    \label{eq:effectiveDirection}
\end{equation}
Therefore, in order to ensure reliable performance for the practical \ac{mMIMO} systems, the effect of $\mathbf{a}_{\text{squint}}$ should be minimized, if not completely eliminated. The transceiver design presented in the subsequent section is dedicated to minimizing the squinting problem taking complexity and energy efficiency issues into account.

% ---------------------------------------------------
\begin{figure}[t!]
  \centering
    \includegraphics[scale=0.41]{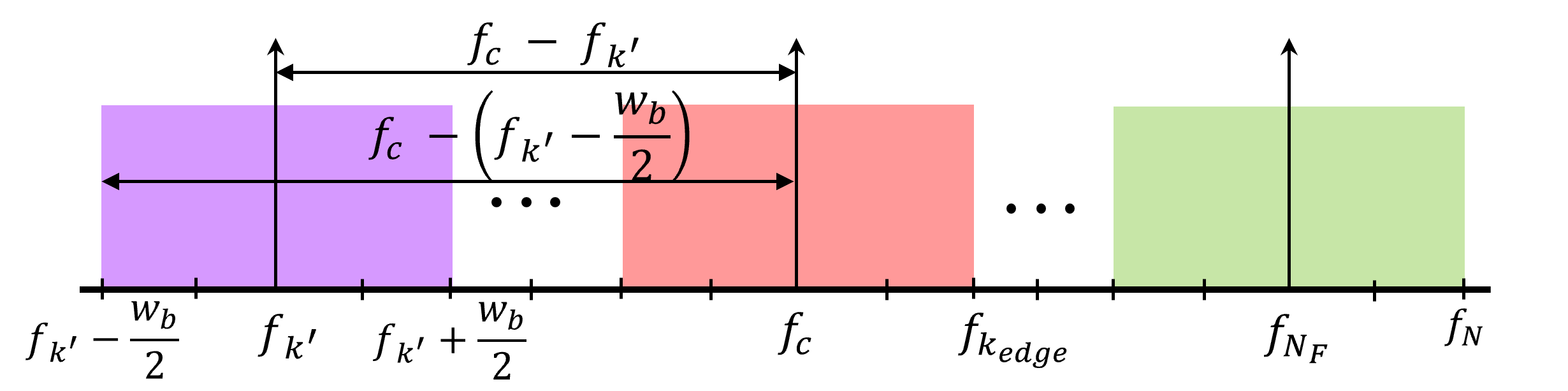}
    \caption{Sub-grouping the subcarriers in the subband filters.}
  \label{fig:subgrouping}
  \vspace{-17pt}
\end{figure}
% ------------------------------------------------

\section{The Proposed LAS System Design} 
\subsection{Transceiver Design} \label{sec:Prposed Design}
This work leverages the unique design of the \ac{LAS} transceiver presented in our earlier work \cite{Murat} that gives the inherent ability to control the signal in the analog domain. The \ac{LAS} design is modified to counteract the beam-squint problem while maintaining all advantages of the conventional LAS design. As shown in Fig. \ref{fig:LAS_system}, the design consists of $N_L$ lenses, each of which equipped with $P$ antenna elements, and $N_L$ \acp{PS}, such that $P\times N_L = M$. Unlike the conventional \ac{LAS} structure, the modified design features analog subband filters for chunking the wideband signal into groups of narrowband signals such that the subcarriers within each group experience approximately similar amount of squinting. It also features the \ac{MPMT} instead of the \ac{SPMT} switching network for relaying all the narrowband groups to their respective antennas elements (based on the degree of their squinting) under the lenses. 

At the transmitter, for one \ac{RF} chain, the \ac{UWB} signal passes through the \ac{PS} network where each \ac{PS} adds a fixed delay to the incoming \ac{RF} signal based on the desired direction $\hat\theta$ \footnote{Note that the multipath index $\ell$ is dropped here and in the subsequent equations just for the sake of notational simplicity.}. As such, the \ac{PS} precoder design is determined by $\mathbf{a}_{\text{ideal}}$ and can be given as
\begin{equation}
    \mathbf{f}_{\textrm{PS}}=\frac{1}{\sqrt{N_L}}\left[ 1, e^{j \frac{2\pi d_L}{\lambda_c} \sin\hat\theta},\cdots, e^{j \frac{2\pi d_L}{\lambda_c} (N_L-1)\sin\hat\theta} \right]^T,
\label{equ:PSprecoder}
\end{equation}
where $d_L$ is the distance between the lenses assumed to be $d_L=P \times d$.
The output of each \ac{PS} is passed through $N_L$ groups of analog subband filters (each group has $N_F$ parallel analog subband filters) to chunk the wideband signal into $N_F$ narrowband signals of bandwidth $W_b$ where $W_b = W/N_F$ and $W$ is the system bandwidth. $N_F$ is decided based on the amount of squint that the system can tolerate. The extreme case is to compensate the squint due to each individual subcarrier, which would require $N_L\times N$ filters for the whole design. To relax the design requirement, it is assumed that the squinting within the quarter of 3 dB beamwidth $\Omega_\text{3dB}$ is tolerable inside each subband, i.e.,
\begin{equation} 
     \bigg|\theta(f_{k_{\text{edge}}}) - \theta(f_{k^{'}}) \bigg| \leq \frac{\Omega_\text{3dB}}{8} ,
    \label{eq:3dB condition}
\end{equation}
where $f_{k^{'}}=\left( (k^{'}-1)-(N_F-1)/2 \right) \Delta f +f_c$ is the center frequency of the filter, with $k^{'}=1,2,\cdots,N_F$, and $f_{k_{\text{edge}}}=f_{k^{'}}\pm W_b/2$ is the carrier frequency of a subcarrier at the edge of $k^{'}$-th subband (see Fig. \ref{fig:subgrouping}). Using \eqref{eq:effectiveDirection}, \eqref{eq:3dB condition} can be simplified to
\begin{equation} 
\begin{split}
  &\bigg| \sin^{-1} \left( \frac{\sin \hat{\theta}}{1+ \frac{\Delta f}{2f_c} (2k^{'}-N_F - \frac{N}{N_F}-1)} \right) -
     \\& \ \sin^{-1} \left( \frac{\sin \hat{\theta}}{1+ \frac{\Delta f}{2f_c} (2k^{'}-N_F-1)} \right) \bigg| \leq \frac{\Omega_\text{3dB}}{8},  \\&     ~~~~~~~~~\textrm {s.t.}~~~ N \mod~ N_F = 0.
\end{split}
\label{eq:N_F optimum}
\end{equation}

\begin{figure}[t]
  \centering
    \includegraphics[scale=0.34]{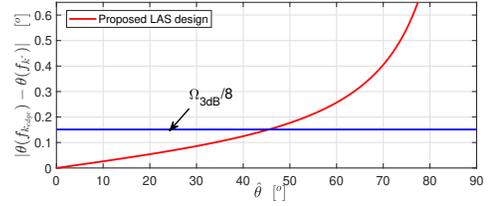}
    \caption{Squinting amount of a given subband for any desired direction within \ac{LAS}'s scanning range is within the tolerable range ($\Omega_\text{3dB}/8$). Here, \eqref{equ:N_F solution} is used to calculate $N_F$, with $W=5$ GHz and $f_c=60$ GHz.}
  \label{fig:phase_behaviour}
  \vspace{-17pt}
\end{figure}

For simplicity, we consider the subband centered at $f_c$ such that $f_{k^{'}} = f_c$ and $k^{'}=(N_F+1)/2$. Consequently, \eqref{eq:N_F optimum} reduces to
\begin{equation} 
\begin{aligned}
    \left| \hat{\theta} -\sin^{-1} \left( \frac{\sin\hat{\theta} }{1+\frac{W}{2f_c N_F}}
      \right) \right|  \leq \frac{\Omega_\text{3dB}}{8} , \\
\begin{split}
     \textrm {s.t.}~~~ & N \mod N_F = 0. 
\end{split}
\end{aligned}
\end{equation}
Considering that the severity of the beam squint increases as $\hat{\theta}$ grows (equation \eqref{eq:squintDelay}), $N_F$ is optimized based on the maximum scanning angle of the \ac{LAS} system (i.e., $\hat{\theta}=\pi/4$ \cite{Murat}) such that it is applicable for all users within the scanning range (See Fig. \ref{fig:phase_behaviour}). This leads to 
\begin{equation}
    N_F \geq \frac{W \sin (\pi/4-\Omega_\text{3dB}/8)}{\sqrt{2}f_c (1-\sqrt{2}\sin (\pi/4-\Omega_\text{3dB}/8))}.
\label{equ:N_F solution}
\end{equation}

Each narrowband signal passes through the switching mechanism using \ac{MPMT} switches modeled by a precoder $F_{\text{LAS}}$ to the lenses for transmission. The \ac{MPMT} switches allow simultaneous activation of $N_F$ (one for each subband) antennas under each lens. $F_{\text{LAS}}$ is given as
\begin{equation} \small
    \mathbf{F}_{\textrm{LAS}} = \begin{bmatrix}
\mathbf{s}^{(0)}    & \mathbf{0}               & \cdots   & \mathbf{0}          \\
\mathbf{0}               & \mathbf{s}^{(1)}     &  & \mathbf{0}          \\
\vdots          &               &  \ddots & \vdots      \\
\mathbf{0}                & \mathbf{0}               & \cdots   & \mathbf{s}^{(N_L-1)} 
\end{bmatrix},
\label{equ:F-LAS}
\end{equation}
where $\mathbf{s}^{(n)}$ is a $P \times 1$ antenna selection vector at $n$-th lens element and $n=0,1,\dots,N_L-1$. The $p$-th element in $\mathbf{s}^{(n)}$ is used to activate the $p$-th antenna element for an incoming ${k^{'}}$-th subband's signal. The activated $p$-th antenna element produces a beam at direction $\theta^{(n)}(p) = \frac{\pi}{4}-\frac{p}{(P-1)}\frac{\pi}{2}$ where $p=0,1,\dots,P-1$ as explained in \cite{Murat}. 

Eventually, the effective \ac{RF} precoder, $\mathbf{f}_\text{RF}$, for the proposed design is $\mathbf{f}_\text{RF} = \mathbf{F}_\text{LAS} \mathbf{f}_\text{PS}$ and its $(Pn+p)$-th element is obtained as $\frac{1}{\sqrt{N_LP}} e^{j2\pi \frac{d}{\lambda_c}Pn\sin\hat{\theta}}\times e^{-j2\pi \frac{d}{\lambda_c}(\frac{(P-1)}{2}-p)\sin\theta^{(n)}}$  as explained in \cite{Murat}.

Let $\theta_{LAS}(k^{'})$ be the $k^{'}$-th subband's effective beam direction given by the proposed \ac{LAS} design with arbitrary values of $\theta^{(n)}$'s across the lenses, then the maximum beam gain $g_\text{max}$ of the design at $k^{'}$-th subband is obtained as
\begin{equation}
\begin{split}
    &g_\text{max}(\mathbf{f}_\text{RF},\theta_\text{LAS}(k^{'}),f_{k^{'}}) =  \mathbf{f}_\text{RF}^H ~ \mathbf{a}(\theta_\text{LAS}(k^{'}),f_{k^{'}}) \\& = \frac{1}{N_LP} \sum_{n=0}^{N_L-1} \sum_{p=0}^{P-1}  e^{-j2\pi \frac{d}{c} (Pn+p)(f_{k^{'}}+f_c) \sin\theta_\text{LAS}(k^{'})} \\& ~~~ \times e^{-j2\pi \frac{d}{c}f_cPn\sin\hat{\theta}} e^{-j2\pi \frac{d}{c}f_c(\frac{(P-1)}{2}-p)\sin\theta^{(n)}}. %\\& = 1
\end{split}
\label{equ:beam_pattern}
\end{equation}
For each subband, the proposed design seeks to find the set $\Phi(k^{'})=\{ \theta^{(1)},\theta^{(2)},\dots,\theta^{(n)},\dots,\theta^{(N_L)}\}$ that gives optimum $\theta_\text{LAS}(k^{'})$, i.e., $\underset{\Phi(k^{'})} \min|\hat{\theta}-\theta_\text{LAS}(k^{'})|$. This can be done by the exhaustive search approach whose complexity for each subband can be quantified as $\mathcal{O}\left(P^{N_L}-k^{'}+1\right)$.

\subsection{Proposed Threshold-based Precoding} \label{sec:Precoder Design}
In order to reduce the complexity of the exhaustive search approach, a threshold-based search mechanism is proposed. In this case, a threshold is specified (based on the desired performance) to allow the selection of a sub-optimal set of antennas over the lenses for a given $k^{'}$-th subband. Once the sub-optimal set that satisfies the specified threshold is found, the search process continues with the next subband. This avoids the searching over all ($P^{N_L}-k^{'}+1$) possible combinations of the antenna under all lenses for a given subband, thereby reducing the complexity. 

In the proposed precoder, it is assumed that, for the middle subband (i.e., $f_{k^{'}}|_{k^{'}= N_F/2}$) the beam deviation within specific threshold (i.e., $\Omega_\textrm{3dB}$ of the desired beam direction) is acceptable. Since different subbands experience different amounts of squint, the threshold is scaled based on $f_{k^{'}}$ with respect to $f_c$. Hence, the search seeks to find a sub-optimal set of antennas that satisfy $|\hat{\theta}-\theta_\text{LAS}(k^{'})|\leq \frac{\Omega_\textrm{3dB}}{2}\times\left(1+\left|\frac{f_{k^{'}}-f_c}{f_c}\right|\right)$ for $k^{'}$-th subband. Since the squint causes beam deviation from the desired direction $\hat{\theta}$, the search process for a given subband begins around $\hat{\theta}$ by setting the initial beam direction under the lenses as $\Phi_\text{init}(k^{'}) =\{\hat{\theta},\hat{\theta},\dots,\hat{\theta}\}$. Since each subband has a different amount of squinting, the combination of $\theta^{(n)}$'s that makes up the optimum set $\Phi(k^{'})$ is unique for each subband. Therefore, such a combination can be removed from the search sets of the next subbands, thereby reducing the searching complexity even further. Note that, with the $\Phi(k^{'})$ initialization approach proposed above, the subband with the least squinting converges faster. As such, the searching process is initialized from the middle to the edge subbands to reduce the size of the search sets at the edge subbands.

The proposed \ac{LAS} design with its precoding are summarized in Algorithm \ref{algorithm:LASdesign}.

% ----------------------------
\begin{algorithm}
\DontPrintSemicolon
  \KwInput{$f_c$, $W$, $M$, $\hat{\theta}$.}
  \KwOutput{$\mathbf{F}^{\textrm{(opt)}}_\textrm{LAS}$.}
  \textbf{Calculate} the optimum number of analog subband filters $N_F$ as specified in \eqref{equ:N_F solution}. \;
  \textbf{Build} the proposed \ac{LAS} design as illustrated in Fig. \ref{fig:LAS_system}.\;
  \textbf{Set} the \ac{PS} precoder as in \eqref{equ:PSprecoder}. \;
  \textbf{Define} all possible sets for $\mathbf{s}_t^{(n)}$ to build a possible $\mathbf{F}_\text{LAS}$. \;
  \textbf{Define} the half-power beamwidth $\Omega_\textrm{3dB}$ of the design. \;
%   \For{$k^{'}$=$(N_F+1)/2,(N_F+1)/2 \pm 1, (N_F+1)/2 \pm 2$, $\dots$}
    \For{$k^{'}=N_F/2, N_F/2 \pm 1, \dots$}
   {  \textbf{Select} the initial set of searching as $\Phi_\text{init}(k^{'})=\{\hat{\theta},\hat{\theta},\dots,\hat{\theta}\}$.  \;
     \textbf{Calculate} $\theta_\text{LAS}(k^{'})$ that satisfies \eqref{equ:beam_pattern}.\; 
 \While {$|\hat{\theta}-\theta_\text{LAS}(k^{'})| > \frac{\Omega_\textrm{3dB}}{2}\times\left(1+\left|\frac{f_{k^{'}}-f_c}{f_c}\right|\right)$}
        { 
        \textbf{Go} to the next set $\Phi(k^{'})$. \;
        \textbf{Repeat} step 8. \;
        % \textbf{Select} one set from the possible antenna selection sets starting from the set that satisfy $\theta^{(n)}\approx \hat{\theta}$ under all lenses. \;
        
        }
    \textbf{Assign} the sub-optimum set to the given subband and remove this set from the possible sets for the next subband. \; 
           }
    \textbf{Build} $\mathbf{F}_\textrm{LAS}$ from the optimum selected sets.\;
\caption{The proposed LAS design and Threshold-based Precoding.}
\label{algorithm:LASdesign}
 \vspace{-1.5pt}
\end{algorithm}
% ----------------------------------

\section{Numerical Results and Discussion}
In this section, performance of the proposed design is analyzed in terms of beam gain, precoder complexity, capacity, and power consumption.  Unless stated otherwise, the numerical values of the used system parameters are $W=5$ GHz, $f_c=60$ GHz, $\hat\theta=\pi/4$, $M=128$, $N_L=4$, $N_F=16$, the number of \ac{RF} chains $N_\text{RF}=1$, $N=2048$, $L_p=1$, and path gain $\tilde\alpha$ follows the complex normal distribution  $\tilde\alpha \sim \mathcal{C} \mathcal{N}(0,1)$.

\begin{figure}[t]
  \centering
    \includegraphics[scale=0.4]{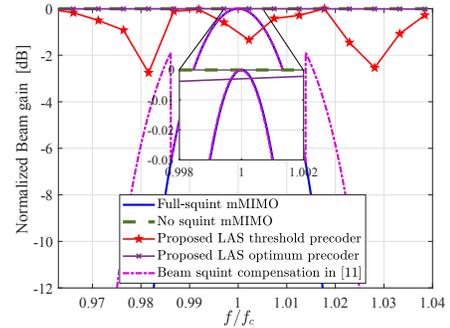}
    \caption{Beam gain vs normalized frequency.}
  \label{fig:beamgain}
  \vspace{-17pt}
\end{figure}

\begin{figure}[t!]
  \centering
    \subfloat[]{\includegraphics[scale=0.39]{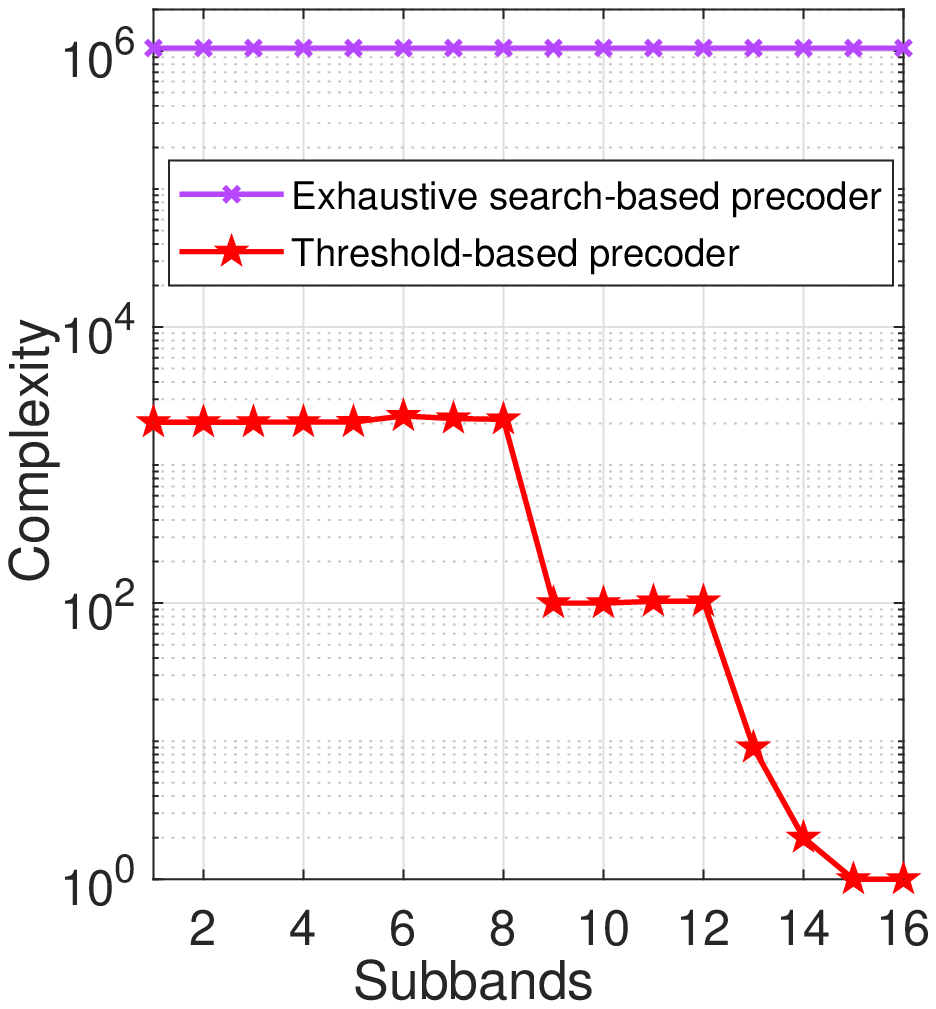}\label{fig:Complexity}}
    \subfloat[]{\includegraphics[scale=0.39]{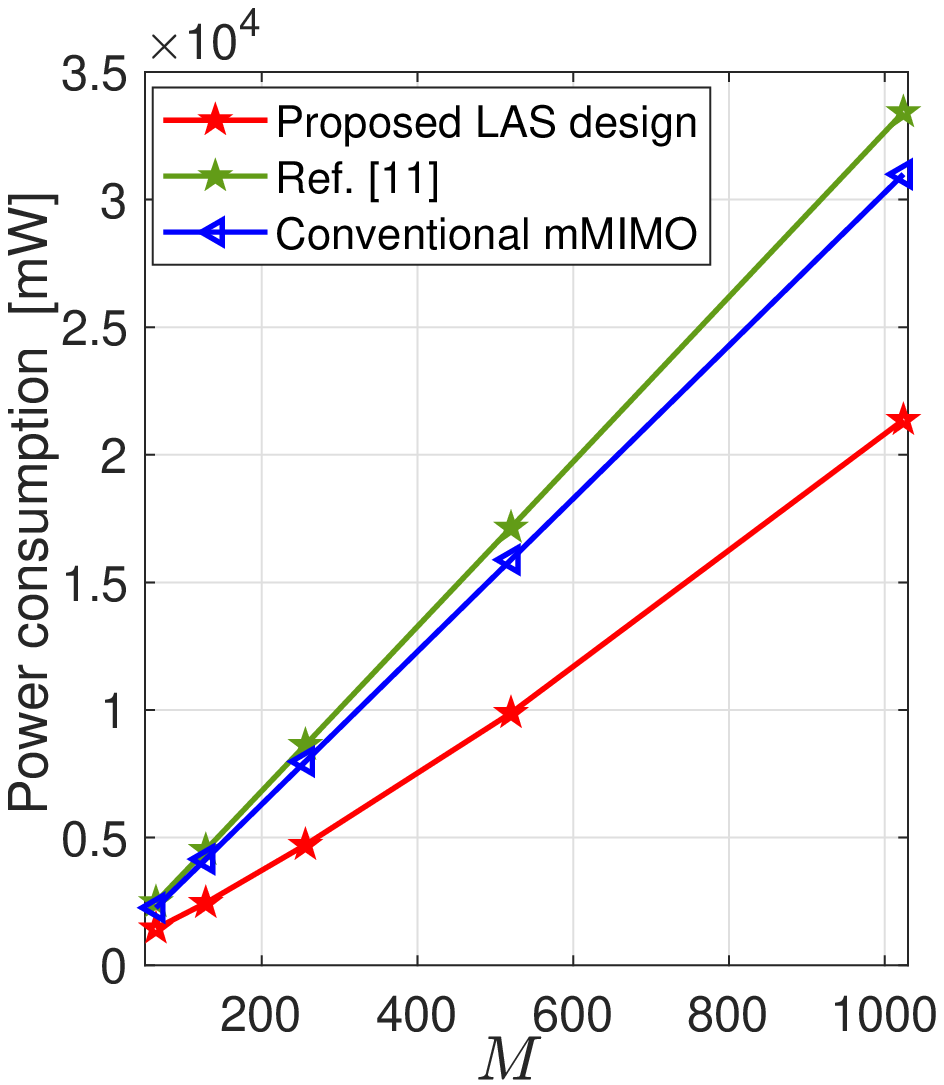}\label{fig:power_consumption}}
    \caption{(a) Complexity analysis, and (b) power consumption vs $M$ analysis.}
\end{figure}

The achievable beam gain with the proposed \ac{LAS} design is given in Fig. \ref{fig:beamgain}, bench-marked with the systems with full and zero beam squint as well as with \cite{sattar2019antenna}. The beam gain provided by the proposed design is analyzed with both full exhaustive search and the proposed threshold-based search approach. While both cases significantly improve the gain (to less than $\approx$3 dB of the ideal performance), the exhaustive search, as expected, performs relatively better, though at the expense of the high complexity as analyzed in Fig. \ref{fig:Complexity}. Fig. \ref{fig:Complexity} shows a huge difference in complexity between the exhaustive search (which is in the order of millions) and the proposed threshold-based precoder (only hundreds to few thousands) at each subband. This is due to the simplified searching mechanism adopted by the proposed precoder as explained in Section \ref{sec:Precoder Design}.

Next, the power consumption of the proposed design is analyzed and compared with other related designs in the literature. The power consumption model of the proposed \ac{LAS} design is divided into four parts, given as
\begin{equation} \small
    P_c = \frac{P_x}{\eta_\textrm{PA}\eta_\textrm{SW}\eta_\textrm{filter}} + N_\text{RF} N_L P_\textrm{PS} + N_\text{RF} P_\textrm{SW} + N_\text{RF} P_\textrm{RF},
\label{equ:power_consumption}
\end{equation}
where $P_x$ is the transmitted power \cite{Murat}, $\eta_\textrm{PA}$, $\eta_\textrm{SW}=10^{-\textrm{IL}_\textrm{SW}/10}$, and $\eta_\textrm{filter}=10^{-\textrm{IL}_\textrm{filter}/10}$ are power efficiencies of the amplifier, switches, and subband filters, respectively, with ${\textrm{IL}}_{\textrm{SW}}$ and ${\textrm{IL}}_{\textrm{filter}}$ being insertion losses of the switches and the filters. $P_\textrm{PS}$, $P_\textrm{SW}$, and $P_\textrm{RF}$ are the power consumption of \ac{PS}, \ac{MPMT} switch, and \ac{RF} chain, respectively. In our analysis, we consider $\eta_\textrm{PA}=0.2$, $\textrm{IL}_\textrm{filter}=1\times N_F$ dB, $P_\textrm{PS}=30$ mW, and $P_\textrm{RF}=220$ mW as used in \cite{Murat}.

Since the values of $P_\textrm{SW}$ are only available for the few ports \ac{MPMT} switches, we model the $N_F$-poles $\times$ $P$-throws \ac{MPMT} switch considered in our design as a series of \ac{SPMT} switches for each pole for power consumption analysis. Therefore, the term $P_\textrm{SW}$ in \eqref{equ:power_consumption} becomes $\log_2(P)N_L L P_\textrm{SW}^{(s)}$, where $P_\textrm{SW}^{(s)} = 10$ mW is the power consumption of an \ac{SPMT} switch \cite{Murat}. The calculated power consumption for different array size is summarized in Fig. \ref{fig:power_consumption}. The performance is compared with that of the conventional \ac{mMIMO} system as analyzed in \cite{Murat}, and with reference \cite{sattar2019antenna} which also presents subband-based beam squinting compensation design. Note that, the power consumption model for the design presented in \cite{sattar2019antenna} follows \eqref{equ:power_consumption} excluding the switching related terms. It is clear from Fig. \ref{fig:power_consumption} that the proposed \ac{LAS} design is the most power-efficient design while profitably compensating for the beam squinting, which makes it suitable for commercial applications of \ac{mMIMO} networks.

\begin{figure}[t!]
  \centering
    \subfloat[$N_F = 16 $]{\includegraphics[scale=0.44]{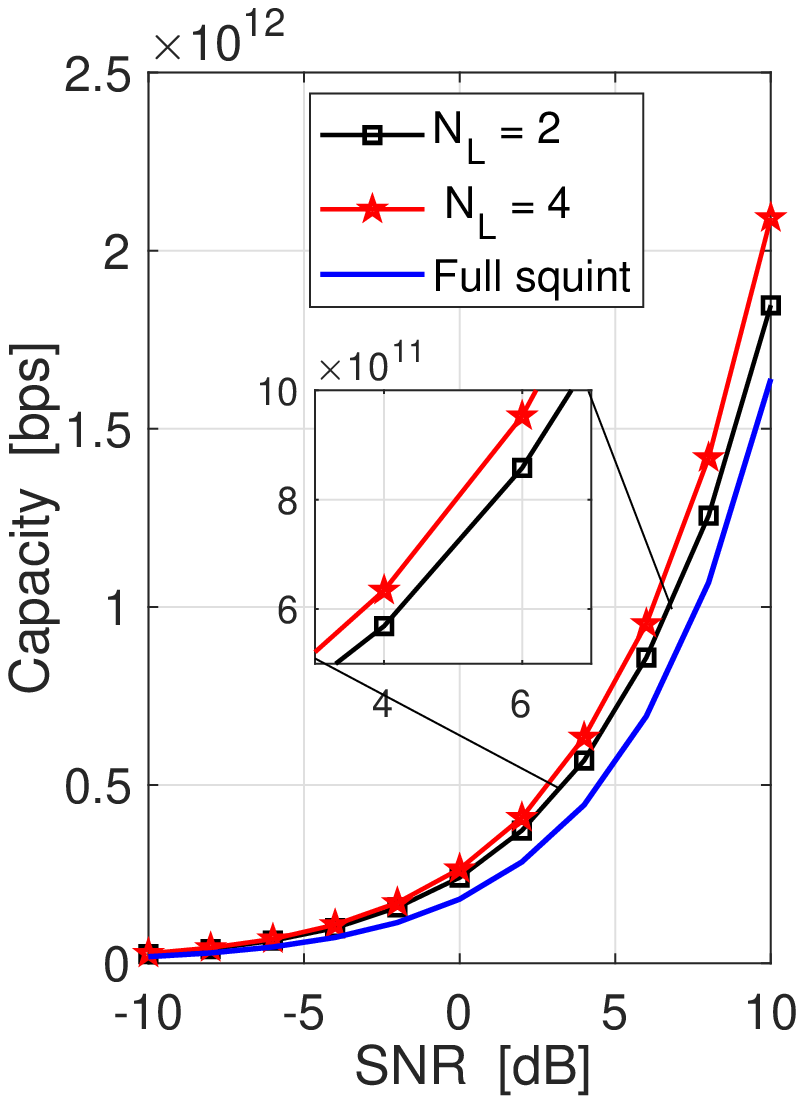}\label{fig:capacityNF16}}
    \subfloat[$N_L = 4 $]{\includegraphics[scale=0.44]{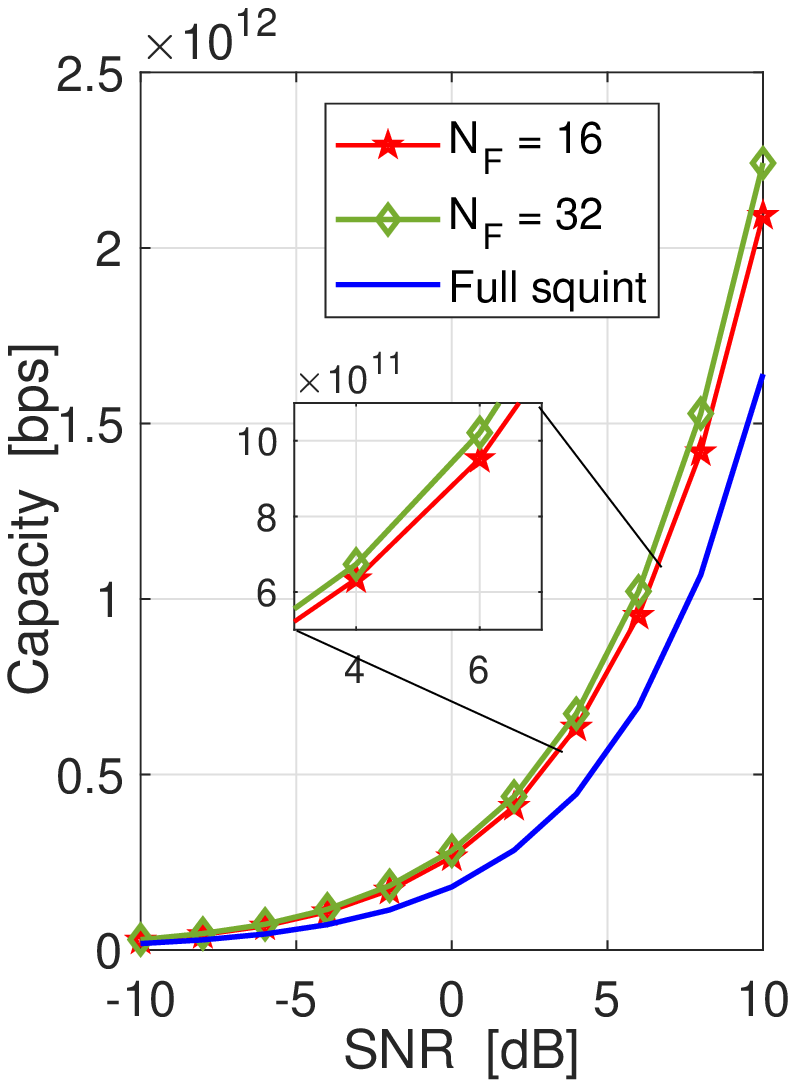} \label{fig:capacityNL4}}
    \caption{Capacity vs SNR analysis.}
  \label{fig:capacity}
  \vspace{-17pt}
\end{figure}

Fig. \ref{fig:capacity} gives some insights regarding the effects of the design parameters $N_F$ and $N_L$ on the system capacity of the proposed design. Regarding the point-to-point link scenario with the single-antenna receiver considered in this work, the system capacity is evaluated as \cite{gherekhloo2020hybrid}
\begin{equation} \small
	 C = W \EX \left\{ \sum_{k=1}^{N}\log_2 \left( 1 + \gamma \mathbf{h}(f_k) \mathbf{f}_\text{RF} \mathbf{f}_\text{RF}^H \mathbf{h}^H(f_k)   \right) \right\} ,
\label{equ:capacity}    
\end{equation}
where $\gamma$ is the SNR and $\EX\{.\}$ is the expectation operation. Fig. \ref{fig:capacityNF16} shows that there is slight capacity improvement as $N_L$ increases. We reckon that this improvement stems from increasing the controllablity of the resultant beam as $N_L$ increases. Fig. \ref{fig:capacityNL4} shows that the system performance improves slightly with the increase in $N_F$, which is plausible due to the fact that as $N_F$ increases, the subband sizes decreases, leading to less residual squint within each subband. However, a large number of filters increases the system's power consumption. Therefore, the choice of $N_F$ in \eqref{equ:N_F solution} should consider the power efficiency issue as well, which is left for future work.

\section {Conclusion}

In this paper, a novel analog transceiver design based on the LAS architecture and subband filters is proposed to compensate for the beam squint effect in UWB mMIMO systems. While the proposed design works well with the traditional exhaustive search-based precoder, a less complex, threshold-based precoding technique is also proposed. Simulation results demonstrate that the proposed LAS-based transceiver design provides good system performance with both the full and threshold-based simplified exhaustive search precoding. Although the proposed design can be implemented with off-the-shelf components, in practice, high-quality narrowband filters should be used to ensure better performance. Research on such filters for higher frequency bands is ongoing \cite{qorvowhitepaper}. Extension of the proposed design to multi-user scenarios with hybrid precoding is left for future work.

\balance

\ifCLASSOPTIONcaptionsoff
  \newpage
\fi


\end{document}